\documentclass[letterpaper, 10pt, conference]{ieeeconf}      % Use this line for a4
 \pdfoutput = 1
\IEEEoverridecommandlockouts                              % This command is only
\usepackage{amsmath}    % need for sub equations
\usepackage{amsfonts}
\usepackage{graphicx}   % need for figures
\usepackage{subcaption}
\usepackage{epsfig} 
\usepackage{color}
\usepackage[normalem]{ulem}
\usepackage{cancel}
\usepackage{amssymb}
\usepackage{color}
\usepackage[ruled,vlined,titlenotnumbered]{algorithm2e} 

\title{\LARGE \bf
Safe Platooning of Unmanned Aerial Vehicles via Reachability}

\author{Mo Chen, Qie Hu, Casey Mackin, Jaime F. Fisac, and Claire J. Tomlin
\thanks{This work is supported in part by NSF under CPS:ActionWebs (CNS-0931843) and CPS:FORCES (CNS1239166), by NASA under grants NNX12AR18A and UCSCMCA-14-022 (UARC), by ONR under grants N00014-12-1-0609, N000141310341 (Embedded Humans MURI), and MIT\_5710002646 (SMARTS MURI), and by AFOSR under grants UPenn-FA9550-10-1-0567 (CHASE MURI) and the SURE project. The research of M. Chen has received funding from the ``NSERC PGS-D'' Program. The research of J.F. Fisac has received funding from the ``la Caixa" Foundation.}
\thanks{All authors are with the Department of Electrical Engineering and Computer Sciences, University of California, Berkeley. \{mochen72,qiehu,cmackin,jfisac,tomlin\}@berkeley.edu}
}

\begin{document}
\maketitle
\thispagestyle{empty}
\pagestyle{empty}

%%%
\begin{abstract}
Recently, there has been immense interest in using unmanned aerial vehicles (UAVs) for civilian operations such as package delivery, firefighting, and fast disaster response. As a result, UAV traffic management systems are needed to support potentially thousands of UAVs flying simultaneously in the airspace, in order to ensure their liveness and safety requirements are met. Hamilton-Jacobi (HJ) reachability is a powerful framework for providing conditions under which these requirements can be met, and for synthesizing the optimal controller for meeting them. However, due to the curse of dimensionality, HJ reachability is only tractable for a small number of vehicles if their set of maneuvers is unrestricted. In this paper, we define a platoon to be a group of UAVs in a single-file formation. We model each vehicle as a hybrid system with modes corresponding to its role in the platoon, and specify the set of allowed maneuvers in each mode to make the analysis tractable. We propose several liveness controllers based on HJ reachability, and wrap a safety controller, also based on HJ reachability, around the liveness controllers. For a single altitude range, our approach guarantees safety for one safety breach; in the unlikely event of multiple safety breaches, safety can be guaranteed over multiple altitude ranges. We demonstrate the satisfaction of liveness and safety requirements through simulations of three common scenarios.
\end{abstract}

% !TEX root = platooning.tex
\section{Introduction}
Unmanned aerial vehicle (UAV) systems have in the past been mainly used for military operations \cite{Tice91}. Recently, however, there has been an immense surge of interest in using UAVs for civil applications through projects such as Amazon Prime Air and Google Project Wing \cite{PrimeAir,ProjectWing,Debusk10}. As a result, government agencies such as the Federal Aviation Administration (FAA) and National Aeronautics and Space Administration (NASA) of the United States are also investigating air traffic control for autonomous vehicles to prevent collisions among potentially numerous UAVs \cite{FAA13}. 

Optimal control and game theory are powerful tools for providing liveness and safety guarantees to controlled dynamical systems, and various formulations \cite{Bokanowski10,Mitchell05,Barron89} have been successfully used to analyze problems involving a small number of vehicles \cite{Fisac15,Chen14,Ding08}. These formulations are based on Hamilton-Jacobi (HJ) reachability, which computes the backwards reachable set, defined as the set of states from which a system is guaranteed to have a control strategy to reach a target set of states. Reachability is powerful because it can be used for synthesizing both controllers that steer the system towards a goal (liveness controllers), and controllers that steer the system away from danger (safety controllers). Furthermore, HJ formulations are flexible in terms of system dynamics, enabling the analysis of nonlinear systems. The power and success of HJ reachability analysis in previous applications is evident, especially since numerical tools are readily available to solve the associated HJ Partial Differential Equation (PDE) \cite{LSToolbox,Osher02,Sethian96}. However, the computation is done on a grid, making the problem complexity scale exponentially with the number of states, and therefore with the number of vehicles. This makes the computation intractable for large numbers of vehicles. 

A considerable body of work has been done on the platooning of vehicles \cite{Kavathekar11}. For example, \cite{McMahon90} investigated the feasibility of vehicle platooning in terms of tracking errors in the presence of disturbances, taking into account complex nonlinear dynamics of each vehicle.  \cite{Hedrick92} explored several control techniques for performing various platoon maneuvers such as lane changes, merge procedures, and split procedures. In \cite{Lygeros98}, the authors modeled vehicles in platoons as hybrid systems, synthesized safety controllers, and analyzed throughput. Finally, reachability analysis was used in \cite{Alam11} to analyze a platoon of two trucks in order to reduce drag by minimizing the following distance while maintaining collision avoidance safety guarantees.

Previous analyses of a large number of vehicles typically do not provide liveness and safety guarantees to the extent that HJ reachability does; however, HJ reachability typically cannot be used to tractably analyze a large number of vehicles. %In the context of HJ reachability, putting vehicles into platoons is desirable because of the additional structure that platoons impose on its members. With additional structure, pairwise safety guarantees of vehicles can be more easily translated into safety guarantees of all the vehicles in the platoon. 
In this paper, we attempt to reconciliate this trade-off by assuming a single-file platoon, which provides structure that allows pairwise safety guarantees from HJ reachability to translate to safety guarantees for the whole platoon. We first propose a hybrid systems model of UAVs in platoons to capture this structure. Then, we show how HJ reachability can be used to synthesize \textit{liveness controllers} that enable vehicles to reach a set of desired states, and wrap \textit{safety controllers} around the liveness controllers to prevent dangerous configurations such as collisions. Finally, we show simulation results of quadrotors forming a platoon, a platoon responding to a malfunctioning member, and a platoon responding to an outside intruder to illustrate the behavior of vehicles in these scenarios and demonstrate the guarantees provided by HJ reachability.

%Motivation:
%\begin{itemize}
%\item Applications of UAVs, potential numbeurs
%\item Importance of safety guarantees
%\item Computation complexity
%\end{itemize}
%
%Related work:
%\begin{itemize}
%\item Platooning references: lack of (?) safety guarantees
%\begin{itemize}\item\textcolor{blue}{Mention PATH papers \& Lygeros thesis}\end{itemize}
%\item HJI, safety guarantees, limited by dimensionality
%\end{itemize}
%
%Summary of results
%\begin{itemize}
%\item Hybrid systems model of vehicles
%\item Reachability guarantees wrapped around existing methods
%\item Reachability offers flexibility in terms of design
%\item Illustrative platoon functions
%\end{itemize}
% Introduction (1p)
%% Motivation
%% Related work
%% Summary of results

% !TEX root = platooning.tex
\section{Problem Formulation \label{sec:formulation}}
\subsection{Vehicle Dynamics}
Consider a UAV whose dynamics are given by
\begin{equation}
\dot{x} = f(x,u)
\end{equation}

\noindent where $x\in\mathbb{R}^n$ represents the state, and $u\in\mathbb{R}^{n_u}$ represents the control action. In this paper, we will assume that each vehicle has a simple kinematics model of a quadrotor:

\begin{equation} \label{eq:dyn}
\begin{aligned}
\dot{p}_x &= v_x, \qquad \dot{p}_y = v_y  \\
\dot{v}_x &= u_x, \qquad \dot{v}_y = u_y, \qquad |u_x|,|u_y| \le u_\text{max}
\end{aligned}
\end{equation}

\noindent where the state (at a fixed altitude) $x=(p_x, v_x, p_y, v_y)\in\mathbb{R}^4$ represents the quadrotor's position in the $x$ direction, its velocity in the $x$ direction, and its position and velocity in the $y$ direction, respectively. For convenience, we will denote the position and velocity $p=(p_x, p_y),v=(v_x,v_y)$, respectively. We will consider a group of $N$ quadrotors $Q_i, i=1\ldots,N$.

In general, the problem of collision avoidance among $N$ vehicles cannot be tractably solved using traditional dynamic programming approaches because the computation complexity of these approaches scales exponentially with the number of vehicles. Thus, in our present work, we will consider the situation where $N$ quadrotors form a platoon. The structure imposed by the platoon enables us to analyze the liveness and safety of the quadrotors in a tractable manner.

\subsection{Relative Dynamics and Augmented Relative Dynamics}
Besides \eqref{eq:dyn}, we will also consider the relative dynamics between two quadrotors $Q_i,Q_j$. These dynamics can be obtained by defining the relative variables

\begin{equation} \label{eq:rel_var}
\begin{aligned}
p_{x,r} &= p_{x,i} - p_{x,j}, \qquad p_{y,r} = p_{y,i} - p_{y,j}\\
v_{x,r} &= v_{x,i} - v_{x,j}, \qquad v_{y,r} = v_{y,i} - v_{y,j}
\end{aligned}
\end{equation}

We treat $Q_i$ as Player 1, the evader who wishes to avoid collision, and we treat $Q_j$ as Player 2, the pursuer, or disturbance, that wishes to cause a collision. In terms of the relative variables given in \eqref{eq:rel_var}, we have 

\begin{equation}
\begin{aligned}
\dot{p}_{x,r}& = v_{x,r}, &\dot{p}_{y,r} &= v_{y,r} \\
\dot{v}_{x,r}& = u_{x,i} - u_{x,j}, &\dot{v}_{y,r} &= u_{y,i} - u_{y,j}\\
\end{aligned}
\end{equation}

We also augment \eqref{eq:rel_var} with the velocity of $Q_i$ to impose a velocity limit when performing the avoidance maneuver.

\begin{equation} \label{eq:rel_dyn_aug}
\begin{aligned}
\dot{p}_{x,r} &= v_{x,r}, &\dot{p}_{y,r} &= v_{y,r} \\
\dot{v}_{x,r} &= u_{x,i} - u_{x,j}, &\dot{v}_{y,r}&= u_{y,i} - u_{y,j}\\
\dot{v}_{x,i} &= u_{x,i}, &\dot{v}_{y,i} &= u_{y,i} \\
\end{aligned}
\end{equation}

\subsection{Quadrotors in a Platoon\label{subsec:platoon_def}}
We consider a platoon of quadrotors to be a group of $M$ quadrotors $Q_{P_1}, \ldots, Q_{P_M}$ in a single-file formation. Not all of the $N$ quadrotors need to be in a platoon: $\{P_j\}_{j=1}^M \subseteq \{i\}_{i=1}^N$. $Q_{P_1}$ is the leader of the platoon, and $Q_{P_2},\ldots,Q_{P_M}$ are the followers. We will assume that the quadrotors in a platoon travel along an air highway, which is defined by as a path inside a pre-defined altitude range. The quadrotors maintain a separation distance of $b$. In order to allow for close proximity of the quadrotors and the ability to resolve multiple simultaneous safety breaches, we assume that in the event of a malfunction, a quadrotor will be able to exit the altitude range of the highway within a duration of $t_\text{internal}=1.5$. Such a requirement may be implemented practically as an emergency landing procedure to which the quadrotors revert when a malfunction is detected. Each quadrotor must be capable of performing a number of essential cooperative maneuvers. In this paper, we consider the following: 
\begin{itemize}
\item safely merging onto an air highway;
\item safely joining a platoon;
\item reacting to a malfunctioning vehicle in the platoon;
\item reacting to an intruder vehicle;
\item following the highway, a curve defined in space at constant altitude, at a specified speed;
\item maintaining a constant relative position and velocity with the leader of a platoon.
\end{itemize}

\subsection{Vehicles as Hybrid Systems}
A UAV in general may be in a number of modes of operations, depending on whether it is part of a platoon, and in the affirmative case, whether it is a leader or a follower. Therefore, it is natural to model vehicles as hybrid systems \cite{Lygeros98,Lygeros12}. In this paper, we restrict the available maneuvers of each quadrotor depending on the mode. We assume that each quadrotor in the airspace has the following modes:

\begin{itemize}
\item Free: Vehicle not in a platoon. Available maneuvers: merge onto a highway, join a platoon on a highway.
\item Leader: Leader of platoon (could be by itself). Available maneuvers: travel along the highway at a pre-specified speed, merge current platoon with a platoon in front, leave the highway.
\item Follower: Vehicle following the platoon leader. Available maneuvers: follow a platoon, create a new platoon.
\item Faulty: Malfunctioned vehicle in a platoon: reverts to default behavior and descends after a duration of $t_\text{internal}$.
\end{itemize}

\begin{figure}
	\centering
	\includegraphics[width=0.5\textwidth]{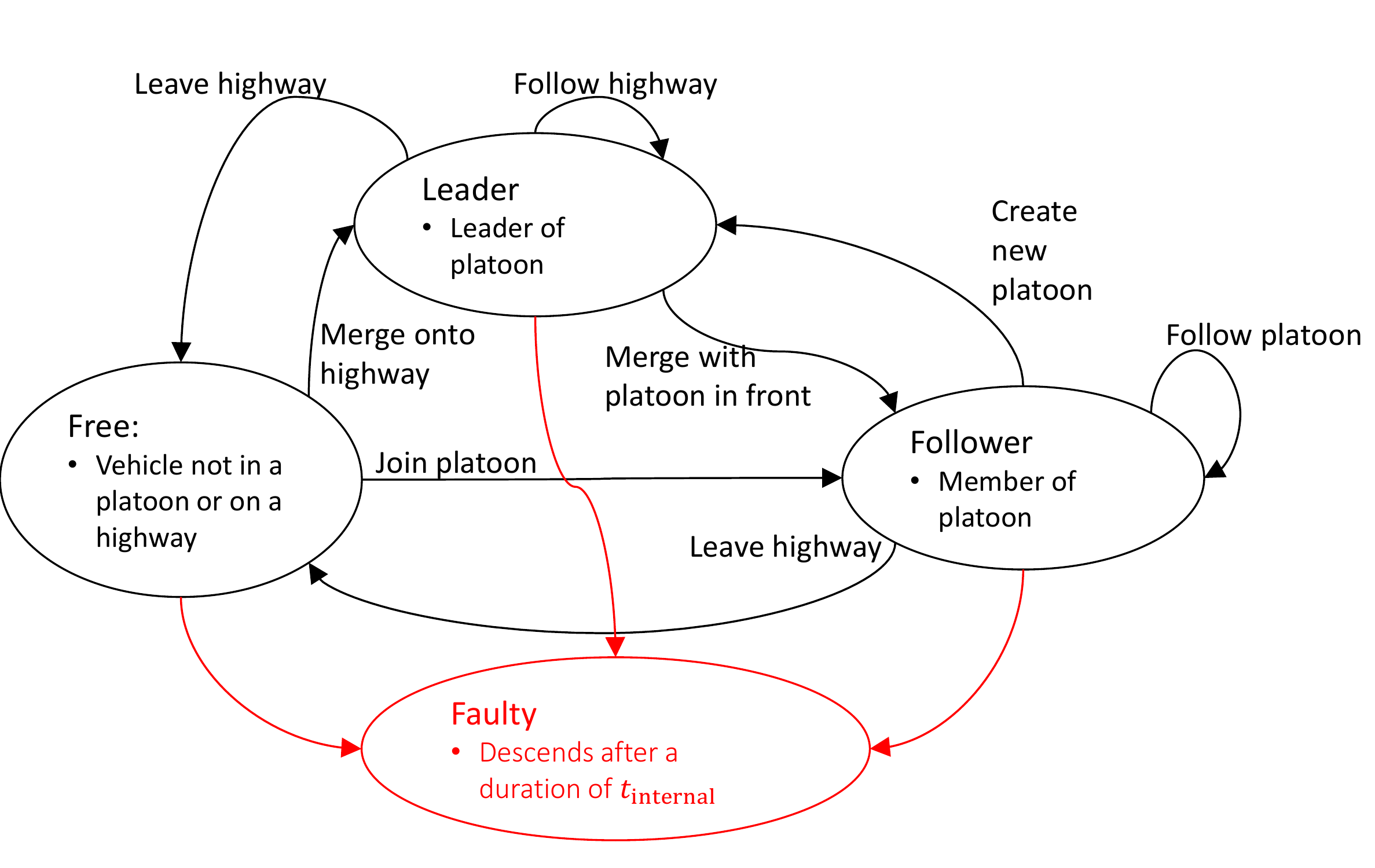}
	\caption{Hybrid modes for vehicles in platoons.}
	\label{fig:vehicleModes}
\end{figure}

The available maneuvers and associated mode transitions are shown in Figure \ref{fig:vehicleModes}.

\subsection{Objectives}
Using the previously-mentioned modeling assumptions, we would like to address the following questions:

\begin{enumerate}
\item How can vehicles effectively form platoons?
\item How can the safety of the vehicles be ensured during normal operation and when there is a malfunctioning vehicle within the platoon?
\item How can the platoon respond to intruders such as unresponsive UAVs, birds, or other aerial objects?
\end{enumerate}

The answers to these questions can be broken down into the maneuvers listed in Section \ref{subsec:platoon_def}. In general, the control strategies of each vehicle have a liveness component, which specifies a set of states towards which the vehicle aims to reach, and a safety component, which specifies a set of states that it must avoid. In this paper, we address both the liveness and safety component using reachability analysis.
% Problem formulation (1.5p)
%% A number of quadrotors forming a platoon in a single file
%% Platoon definition and functions

% !TEX root = platooning.tex
\section{Hamilton-Jacobi Reachability  \label{sec:HJI}}
\subsection{General Framework}
Consider a differential game between two players described by the system
\begin{equation} \label{eq:dyn}
\dot{x} = f(x, u_1, u_2), \text{for almost every }t\in [-T,0]
\end{equation}

\noindent where $x\in\mathbb{R}^n$ is the system state, $u_1\in \mathcal{U}_1$ is the control of Player 1, and $u_2\in\mathcal{U}_2$ is the control of Player 2. We assume $f:\mathbb{R}^n\times \mathcal{U}_1 \times \mathcal{U}_2 \rightarrow \mathbb{R}^n$ is uniformly continuous, bounded, and Lipschitz continuous in $x$ for fixed $u_1,u_2$, and the control functions $u_1(\cdot)\in\mathbb{U}_1,u_2(\cdot)\in\mathbb{U}_2$ are drawn from the set of measurable functions. Player 2 is allowed to use nonanticipative strategies \cite{Evans84,Varaiya67} $\gamma$, defined by

\begin{equation}
\begin{aligned}
\gamma &\in \Gamma := \{\mathcal{N}: \mathbb{U}_1 \rightarrow \mathbb{U}_2 \mid  u_1(r) = \hat{u}_1(r) \\
&\text{for almost every } r\in[t,s] \Rightarrow \mathcal{N}[u_1](r) \\
&= \mathcal{N}[\hat{u}_1](r) \text{ for almost every } r\in[t,s]\}
\end{aligned}
\end{equation}

In our differential game, the goal of Player 2 is to drive the system into some target set $\mathcal{L}$, and the goal of Player 1 is to drive the system away from it. The set $\mathcal{L}$ is represented as the zero sublevel set of a bounded, Lipschitz continuous function $l:\mathbb{R}^n\rightarrow\mathbb{R}$. We call $l(\cdot)$ the \textit{implicit surface function} representing the set $\mathcal{L}=\{x\in\mathbb{R}^n \mid l(x)\le 0\}$.

Given the dynamics \eqref{eq:dyn} and the target set $\mathcal{L}$, we would like to compute the backwards reachable set, $\mathcal{V}(t)$:

\begin{equation}
\begin{aligned}
\mathcal{V}(t) &:= \{x\in\mathbb{R}^n \mid \exists \gamma\in\Gamma \text{ such that } \forall u_1(\cdot)\in\mathbb{U}_1, \\
&\exists s \in [t,0], \xi_f(s; t, x, u_1(\cdot), \gamma[u_1](\cdot)) \in \mathcal{L} \}
\end{aligned}
\end{equation}
where $\xi_f$ is the trajectory of the system satisfying initial conditions $\xi_f(t; x, t, u_1(\cdot), u_2(\cdot))=x$ and the following differential equation almost everywhere on $[-t, 0]$:
\begin{equation}
\begin{aligned}
\frac{d}{ds}&\xi_f(s; x, t, u_1(\cdot), u_2(\cdot)) \\
&= f(\xi_f(s; x, t, u_1(\cdot), u_2(\cdot)), u_1(s), u_2(s))
\end{aligned}
\end{equation}

For this paper, we use the HJ formulation in \cite{Mitchell05}, which has shown that the backwards reachable set $\mathcal{V}(t)$ can be obtained as the zero sublevel set of the viscosity solution \cite{Crandall84} $V(t,x)$ of the following terminal value Hamilton-Jacobi-Isaacs (HJI) PDE:

\begin{equation} \label{eq:HJIPDE}
\begin{aligned}
D_t &V +\min \{0, \max_{u_1\in\mathcal{U}_1} \min_{u_2\in\mathcal{U}_2} D_x V \cdot f(x,u_1,u_2) \} = 0, \\
&V(0,x) = l(x)
\end{aligned}
\end{equation}

\noindent from which we obtain $\mathcal{V}(t) = \{x\in\mathbb{R}^n \mid V(t,x)\le 0\}$. From the solution $V(t,x)$, we can also obtain the optimal controls for both players via the following:

\begin{equation} \label{eq:HJI_ctrl_syn}
\begin{aligned}
u_1^*(t,x) &= \arg \max_{u_1\in\mathcal{U}_1} \min_{u_2\in\mathcal U_2} D_x V(t,x) \cdot f(x,u_1,u_2)\\
u_2^*(t,x) &= \arg \min_{u_2\in\mathcal{U}_2} D_x V(t,x) \cdot f(x,u_1^*,u_2)
\end{aligned}
\end{equation}

In the special case where there is only one player (Player 2 does not exist), we obtain an optimal control problem for a system with dynamics

\begin{equation} \label{eq:dyn_d}
\dot{x} = f(x, u), t\in [-T,0], u\in\mathcal U.
\end{equation}

The reachable set in this case would be given by the Hamilton-Jacobi-Bellman (HJB) PDE

\begin{equation} \label{eq:HJBPDE}
\begin{aligned}
D_t V(t,x) + \min \{0, \min_{u\in\mathcal{U}} D_x V(t,x) \cdot f(x,u)\} &= 0 \\
V(0,x) = l(x)&
\end{aligned}
\end{equation}

\noindent where the optimal control is given by

\begin{equation} \label{eq:HJB_ctrl_syn}
u^*(t,x) = \arg \min_{u\in\mathcal{U}} D_x V(t,x) \cdot f(x,u)
\end{equation}

For our application, we will use a several decoupled system models and utilize the decoupled HJ formulation in \cite{Chen15}, which enables real time 4D reachable set computations and tractable 6D reachable set computations.

\section{Liveness Controllers \label{sec:liveness}}
\subsection{Merging onto a Highway \label{subsec:highway_merge}}
We model the merging of a vehicle onto an air highway as a path planning problem, where we specify a target position and velocity along the highway. Thus, a vehicle would aim to drive the system \eqref{eq:dyn} to a state $\bar{x}_H=(\bar{p}_x, \bar{v}_x, \bar{p}_y, \bar{v}_y)$, or a small range of states defined by the set

\begin{equation}
\begin{aligned}
\mathcal{L}_H = \{x: |p_x-\bar{p}_x|\le r_{p_x}, |v_x-\bar{v}_x|\le r_{v_x}, \\
|p_y - \bar{p}_y| \le r_{p_y}, |v_y - \bar{v}_y|\le r_{v_y} \}.
\end{aligned}
\end{equation}

In this reachability problem, $\mathcal{L}_H$ is the target set, represented by the zero sublevel set of the function $l_H(x)$, which specifies the terminal condition of the HJB PDE to be solved. The solution we obtain, $V_H(t,x)$, is the implicit surface function representing the reachable set $\mathcal V_H(t)$; $V_H(-T,x)\le 0$, then, specifies the reachable set $\mathcal{V}_H(T)$, the set of states from which the system can be driven to the target $\mathcal{L}_H$ within a duration of $T$. This gives the algorithm for merging onto the highway:

\begin{enumerate}
\item Move towards $\bar{x}_H$ in a straight line until $V_H(-T,x)\le 0$. This simple controller is chosen heuristically.
\item Apply the optimal control extracted from $V_H(-T,x)$ according to \eqref{eq:HJB_ctrl_syn} until $\mathcal{L}_H$ is reached.
\end{enumerate}

\subsection{Merging into a Platoon \label{subsec:platoon_merge}}
We again pose the merging of a vehicle into a platoon on an air highway as a reachability problem. Here, we would like quadrotor $Q_i$ to merge onto the highway and follow another vehicle $Q_j$ in a platoon. Thus, we would like to drive the system given by \eqref{eq:rel_dyn_aug} to a specific $\bar{x}_P = (\bar{p}_{x,r}, \bar{v}_{x,r}, \bar{p}_{y,r}, \bar{v}_{y,r})$, or a small range of relative states defined by the set

\begin{equation}
\begin{aligned}
\mathcal{L}_P = \{x: |p_{x,r}-\bar{p}_{x,r}|\le r_{p_x}, |v_{x,r}-\bar{v}_{x,r}|\le r_{v_x}, \\
|p_{y,r} - \bar{p}_{y,r}| \le r_{p_y}, |v_{y,r} - \bar{v}_{y,r}|\le r_{v_y} \}
\end{aligned}
\end{equation}

The target set $\mathcal{L}_P$ is represented by the implicit surface function $l_P(x)$, which specifies the terminal condition of the HJI PDE \eqref{eq:HJIPDE}. The zero sublevel set of the solution to \eqref{eq:HJIPDE}, $V_P(-T,x)$, gives us the set of relative states from which $Q_i$ can reach the target and join the platoon following $Q_j$ within a duration of $T$. We assume that $Q_j$ moves along the highway at constant speed, so that $u_j(t)$ = 0. The following is a suitable algorithm for a vehicle merging onto a highway and joining a platoon to follow $Q_j$:

\begin{enumerate}
\item Move towards $\bar{x}_P$ in a straight line until $V_P(-T,x)\le 0$.
\item Apply the optimal control extracted from $V_P(-T,x)$ according to \eqref{eq:HJI_ctrl_syn} until $\mathcal{L}_P$ is reached.
\end{enumerate}

\subsection{Other Quadrotor Maneuvers}
For the simpler maneuvers of traveling along a highway and following a platoon, we resort to simpler controllers described below.

\subsubsection{Traveling along a highway} \label{sec:travel_hwy}
We use a model-predictive controller (MPC) for traveling along a highway at a pre-specified speed. Here, a leader quadrotor tracks a constant-altitude path, defined as a curve $\bar{p}(s)$ parametrized by $s\in[0,1]$ in $p=(p_x, p_y)$ space (position space), while maintaining a velocity $\bar{v}(s)$ that corresponds to constant speed in the direction of the highway.

\subsubsection{Following a Platoon} \label{sec:follow_platoon}
Follower vehicles use a feedback control law tracking a nominal position and velocity in the platoon, with an additional feed-forward term given by the leader's acceleration input.

The $i$-th member of the platoon, $Q_{P_i}$, is expected to track a relative position in the platoon $r^i = (r_x^i,r_y^i)$ with respect to the leader's position $p_{P_1}$, and the leader's velocity $v_{P_1}$ at all times. The resulting control law has the form:
\begin{equation}\label{eq:follow}
u^i(t) = k_p \big[p_{P_1}(t) + r^i(t) - p^i(t) \big] + k_v\big[v_{P_1}(t) - v^i(t)\big] + u_{P_1}(t)
\end{equation}
for some $k_p,k_v>0$. A simple rule for determining $r^i(t)$ in a single-file platoon is given for $Q_{P_i}$ as:
\begin{equation}\label{eq:nominal_pos}
r^i(t) = - (i-1) b \frac{v_{P_1}}{\|v_{P_1}\|_2}
\end{equation}
where $b$ is the spacing between vehicles along the platoon. and $\frac{v_{P_1}}{\|v_{P_1}\|_2}$ is the platoon leader's direction of travel.

\section{Safety Controllers \label{sec:safety}}
\subsection{Wrapping Reachability Around Existing Controllers}
A quadrotor can use a liveness controller when it is not in any danger of collision with other quadrotors or obstacles. If the quadrotor could potentially be involved in a collision within the next short period of time, it must switch to a safety controller. In this section, we will demonstrate how HJ reachability can be used to both detect imminent danger and synthesize a controller that guarantees safety within a specified time horizon. For our safety analysis, we will use the model in \eqref{eq:rel_dyn_aug}.

We begin by defining the target set $\mathcal{L}_S$, which characterizes configurations in relative coordinates for which vehicles $Q_i,Q_j$ are considered to be in collision:

\begin{equation}
\begin{aligned}
\mathcal{L}_S = \{x: &|p_{x,r}|, |p_{y,r}|\le d \vee |v_{x,i}| \ge v_\text{max} \vee |v_{y,i}| \ge v_\text{max} \}
\end{aligned}
\end{equation}

With this definition, $Q_i$ is considered to be unsafe if $Q_i$ and $Q_j$ are within a distance $d$ in both $x$ and $y$ directions simultaneously, or if $Q_i$ has exceeded some maximum speed $v_\text{max}$ in either $x$ or $y$ direction. For illustration purposes, we choose $d=2$ meters, and $v_\text{max}= 5$ m/s.

We can now define the implicit surface function $l_S(x)$ corresponding to $\mathcal{L}_S$, and solve the HJI PDE \eqref{eq:HJIPDE} using $l_S(x)$ as the terminal condition. As before, the zero sublevel set of the solution $V_S(t,x)$ specifies the reachable set $\mathcal{V}_S(t)$, which characterizes the states in the augmented relative coordinates, as defined in \eqref{eq:rel_dyn_aug}, from which $Q_i$ \textit{cannot} avoid $\mathcal{L}_S$ for a time period of $t$, if $Q_j$ uses the worst case control. To avoid collisions, $Q_i$ must apply the safety controller according to \eqref{eq:HJI_ctrl_syn} on the boundary of the reachable set in order to avoid going into the reachable set. The following algorithm wraps our safety controller around liveness controllers:

\begin{enumerate}
\item For a specified time horizon $t$, evaluate$V_S(-t,x_i-x_j)$ for all $j\in \mathcal{Q}(i)$.

$\mathcal{Q}(i)$ is the set of quadrotors with which quadrotor $i$ checks safety against. We discuss $\mathcal{Q}(i)$ in Section \ref{subsec:safety_guarantees}.

\item Use the safety or liveness controller depending on the values $V_S(-t,x_i-x_j),j\in \mathcal{Q}(i)$: 

If $\exists j\in \mathcal{Q}(i),V_S(-t,x_i-x_j)\le 0$, then $Q_i,Q_j$ are in potential conflict, and $Q_i$ must use a safety controller; otherwise $Q_i$ uses a liveness controller.
\end{enumerate}

\subsection{Platoon Safety Guarantees \label{subsec:safety_guarantees}}
Under normal operations in a single platoon, each follower quadrotor $Q_{P_i},i>1$ checks whether it is in the safety reachable set with respect to $Q_{P_{i-1}}$ and $Q_{P_{i+1}}$. So $\mathcal{Q}(i) = \{P_{i+1}, P_{i-1}\}$ for $i=P_2,\ldots,P_{N-1}$. Assuming there are no nearby quadrotors outside of the platoon, the platoon leader $Q_{P_1}$ checks safety against $Q_{P_2}$, and the platoon trailer $Q_{P_N}$ checks safety against $Q_{P_{N-1}}$. So $\mathcal{Q}(P_1)=\{P_2\}, \mathcal{Q}(P_N)=\{P_{N-1}\}$. No pair of quadrotors should be in an unsafe configuration if the liveness controllers are well-designed. Occasionally, a quadrotor $Q_k$ may behave unexpectedly due to faults, which may lead to an unsafe configuration.

With our choice of $\mathcal{Q}(i)$ and the assumption that the platoon is in a single-file formation, some quadrotor $Q_i$ would get into an unsafe configuration with $Q_k$, where $Q_k$ is likely to be the quadrotor in front or behind of $Q_i$. In this case, a ``safety breach" occurs. Our synthesis of the safety controller guarantees that between every pair of quadrotors $Q_i,Q_k$, as long as $V_S(-t,x_i- x_k)>0$, $\exists u_i$ to keep $Q_i$ from colliding with $Q_k$ for a desired time horizon $t$, despite the worst case (an adversarial) control from $Q_k$. Therefore, as long as the number of ``safety breaches" is at most one, $Q_i$ can simply use the optimal control $u_i$ to avoid collision with $Q_k$ for the time horizon of $t$. Since by assumption, vehicles in platoons are able to exit the current altitude range within a duration of $t_\text{internal}$, if we choose $t=t_\text{internal}$, the safety breach would always end before any collision can occur. 

Within a duration of $t_\text{internal}$, there is a small chance that additional safety breaches may occur. However, as long as the total number of safety breaches does not exceed the number of affected quadrotors, collision avoidance of all the quadrotors can be guaranteed for the duration $t_\text{internal}$. However, as our simulation results show, putting quadrotors in single-file platoons makes the likelihood of multiple safety breaches low during a quadrotor malfunction and during the presence of one intruder vehicle. 

In the event that multiple safety breaches occur for some of the quadrotors due to a malfunctioning quadrotor within the platoon or an intruding quadrotor outside of the platoon, those quadrotors with more than one safety breach still have the option of exiting the highway altitude range in order to avoid collisions. Every extra altitude range reduces the number of simultaneous safety breaches by $1$, so $K$ simultaneous safety breaches can be resolved using $K-1$ different altitude ranges. 

Given that quadrotors within a platoon are safe with respect to each other, each platoon can be treated as a single vehicle, and perform collision avoidance with other platoons. By treating each platoon as a single unit, we reduce the number of individual quadrotors that need to check for safety against each other, reducing overall computation burden.
% Solution methodology (1.5-2p)
%% Hybrid model of vehicles
%% Reachability
%% Functions of platoons: following, joining, splitting

% \input{analysis.tex}
%% Information pattern, stability (0.5p)

% !TEX root = platooning.tex
\section{Scenario Case Study \label{sec:scenarios}}
In this section, we consider several situations that quadrotors in a platoon on an air highway may commonly encounter, and show via simulations the behaviors that emerge from the controllers we defined in Sections \ref{sec:liveness} and \ref{sec:safety}.

\subsection{Forming a Platoon}
We first consider the scenario in which some quadrotors are trying to merge onto an initially unoccupied highway. In order to do this, each quadrotor first checks for safety with respect to the other quadrotors, and uses the safety controller if necessary, according to Section \ref{sec:safety}. Otherwise, the quadrotor uses the liveness controller described in Section \ref{sec:liveness}. 

For the simulation example, the highway is specified by the line $p_y = 0.5p_x$, the point of entry on the highway is chosen to be $(\bar{p}_x, \bar{p}_y) = (4,2)$, and the target velocity is such that the quadrotors travel at a speed $\bar{v}=3$ along the direction of the highway. This forms the target state $\bar{x}_H=(\bar{p}_x, \bar{v}_x, \bar{p}_y, \bar{v}_y)$, from which we define the target set $\mathcal{L}_H$ as in Section \ref{subsec:highway_merge}.

The first quadrotor that completes merging onto the empty highway creates a platoon and becomes its leader, while subsequent quadrotors form a platoon behind the leader in a pre-specified order according to the process described in Section \ref{subsec:platoon_merge}. Here, we choose $(\bar{p}_{x,r}, \bar{p}_{y,r})$ to be a distance $b$ behind the last quadrotor in the platoon, and $(\bar{v}_{x,r}, \bar{v}_{y,r}) = (0,0)$. This gives us the target set $\mathcal{L}_P$.

\begin{figure}
	\centering
	\includegraphics[width=0.35\textwidth]{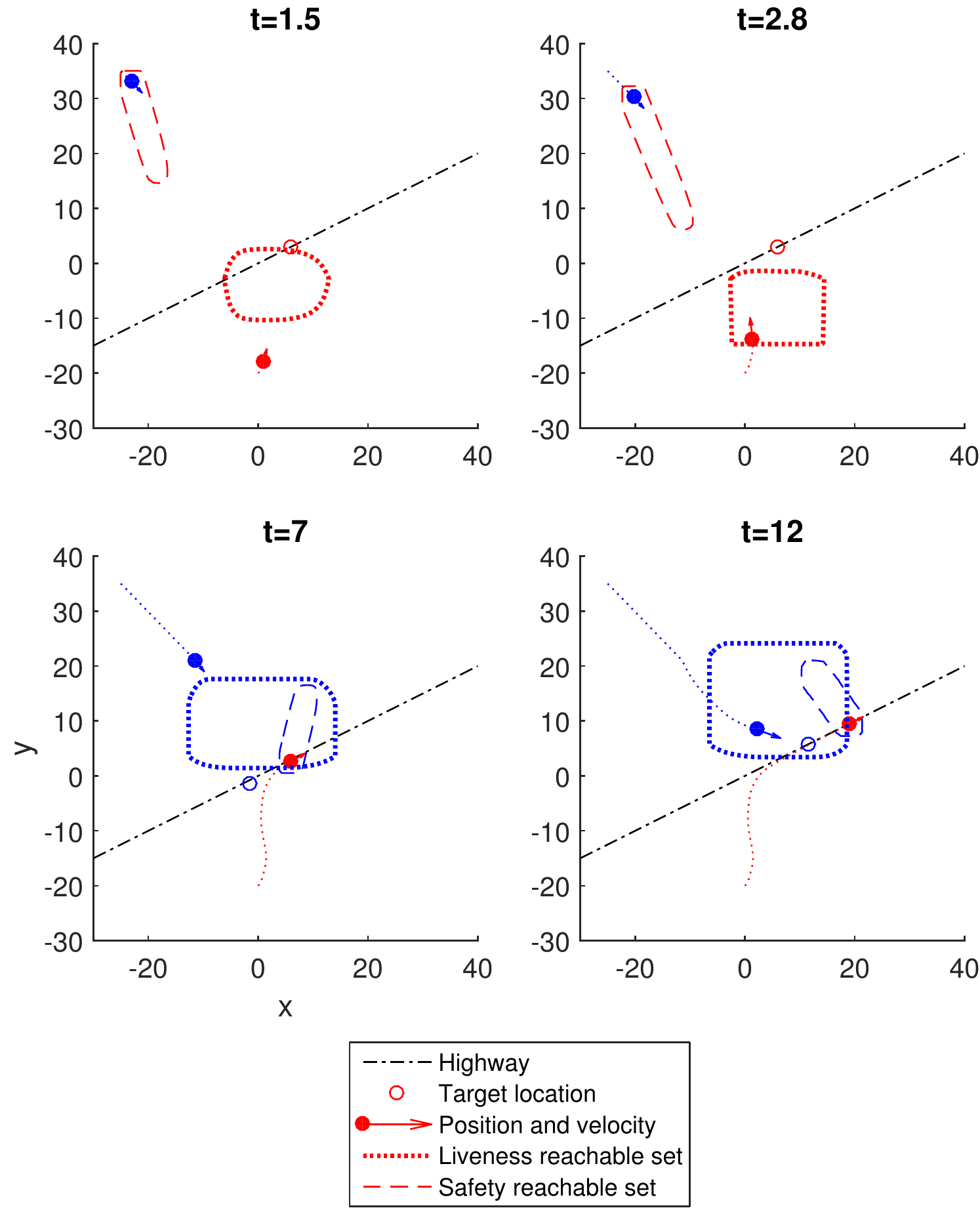}
	\caption{Reachable sets used to merge onto a highway to form a platoon (top subplots) and to join a platoon on the highway (bottom subplots).}
	\label{fig:normal2}
\end{figure}

\begin{figure}
	\centering
	\includegraphics[width=0.35\textwidth]{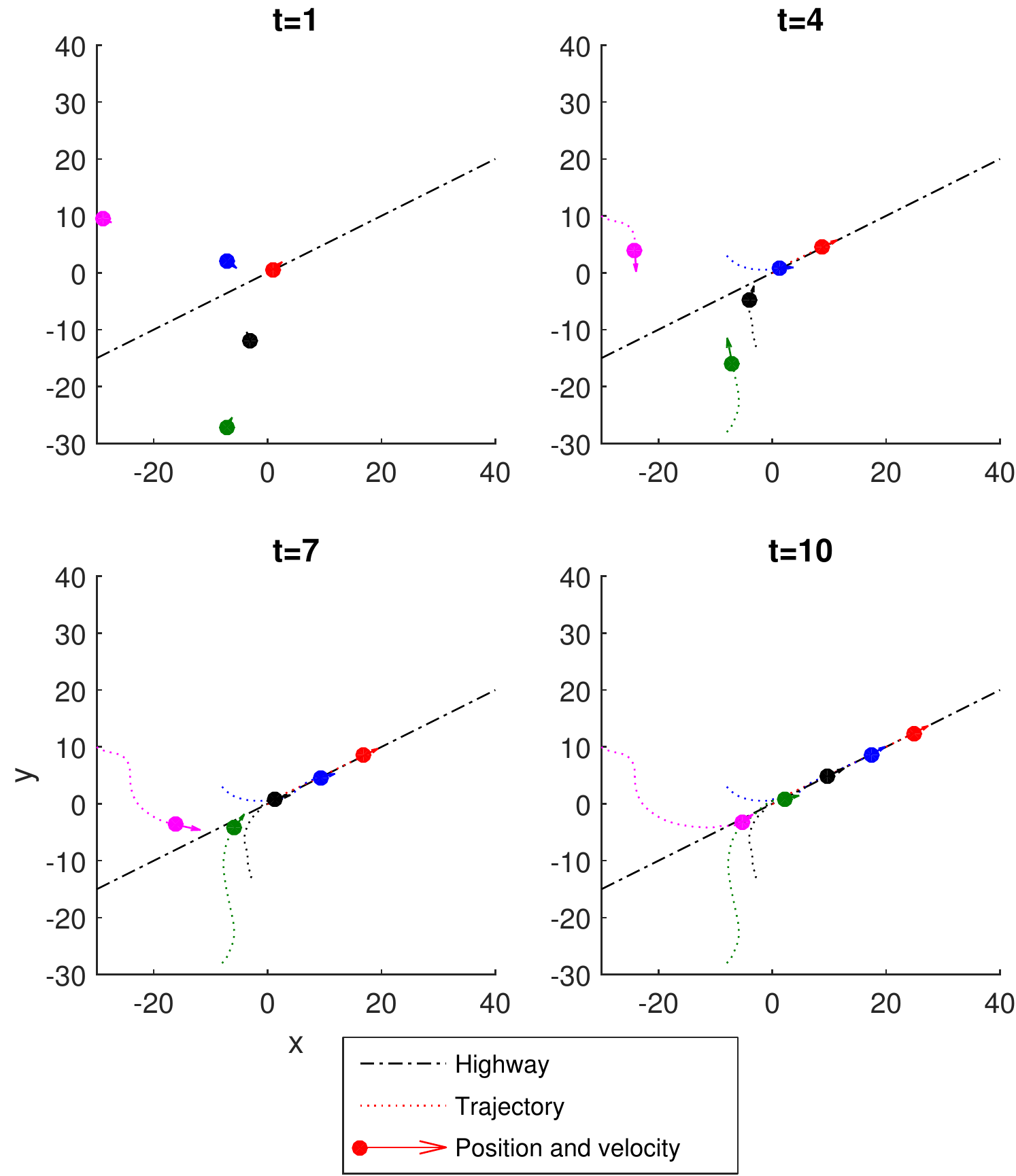}
	\caption{Five quadrotors merging onto a highway.}
	\label{fig:normal5}
\end{figure}

Figures \ref{fig:normal2} and \ref{fig:normal5} show the simulation results. Since the liveness reachable sets are in 4D and the safety reachable sets are in 6D, we compute and plot their 2D slices based on the quadrotors' velocities and relative velocities. 

Figure \ref{fig:normal2} illustrates the use of liveness and safety reachable sets using just two quadrotors to reduce visual clutter. The first quadrotor $Q_1$ (red disk) first travels in a straight line towards the highway merging point $\bar{x}$ (red circle) at $t=1.5$, because it is not yet in the liveness reachable set for merging onto the highway (red dotted boundary). When it is within the liveness reachable set boundary at $t=2.8$, it is ``locked-in" to the target state $\bar{x}_H$, and follows the optimal control in \eqref{eq:HJB_ctrl_syn} to $\bar{x}_H$. During the entire time, $Q_1$ checks whether it may collide with $Q_2$ within a time horizon of $t_\text{external}$; we chose $t_\text{external}=3>t_\text{internal}$. 

After $Q_1$ has reached $\bar{x}_H$, it forms a platoon, becomes the platoon leader, and continues to travel along the highway. $Q_2$ (blue disk), at $t=7$, begins joining the platoon behind $Q_1$, by moving towards the target $\bar{x}_P$ relative to the position of $Q_1$. When $Q_2$ moves inside the liveness reachable set boundary for joining the platoon (blue dotted boundary), it is ``locked-in" to the target relative state $\bar{x}_P$, and begins following the optimal control in \eqref{eq:HJI_ctrl_syn} towards the target whenever it is outside the safety reachable set (blue dashed boundary).

Figure \ref{fig:normal5} shows the behavior of all 5 quadrotors which eventually form a platoon and travel along the highway together. The liveness controllers allow the quadrotors to optimally and smoothly enter the highway and join platoons, while the safety controllers prevent collisions from occurring.

\subsection{Malfunctioning Vehicle in Platoon}
We now consider a scenario where a quadrotor in a platoon of five malfunctions while the platoon is traveling along a highway. To best demonstrate the behavior of the other quadrotors in the platoon, this simulation assumes that $Q_{P_3}$, the middle quadrotor, malfunctions and reverses direction. When this happens, all of the other quadrotors in the platoon begin checking safety against it. In addition, $Q_{P_3}$ is removed from the platoon, causing the other quadrotors to treat it as an intruder.  Trailing quadrotors must leave the highway to avoid colliding with the faulty quadrotor. 

\begin{figure}
	\center
		\includegraphics[width=0.4\textwidth]{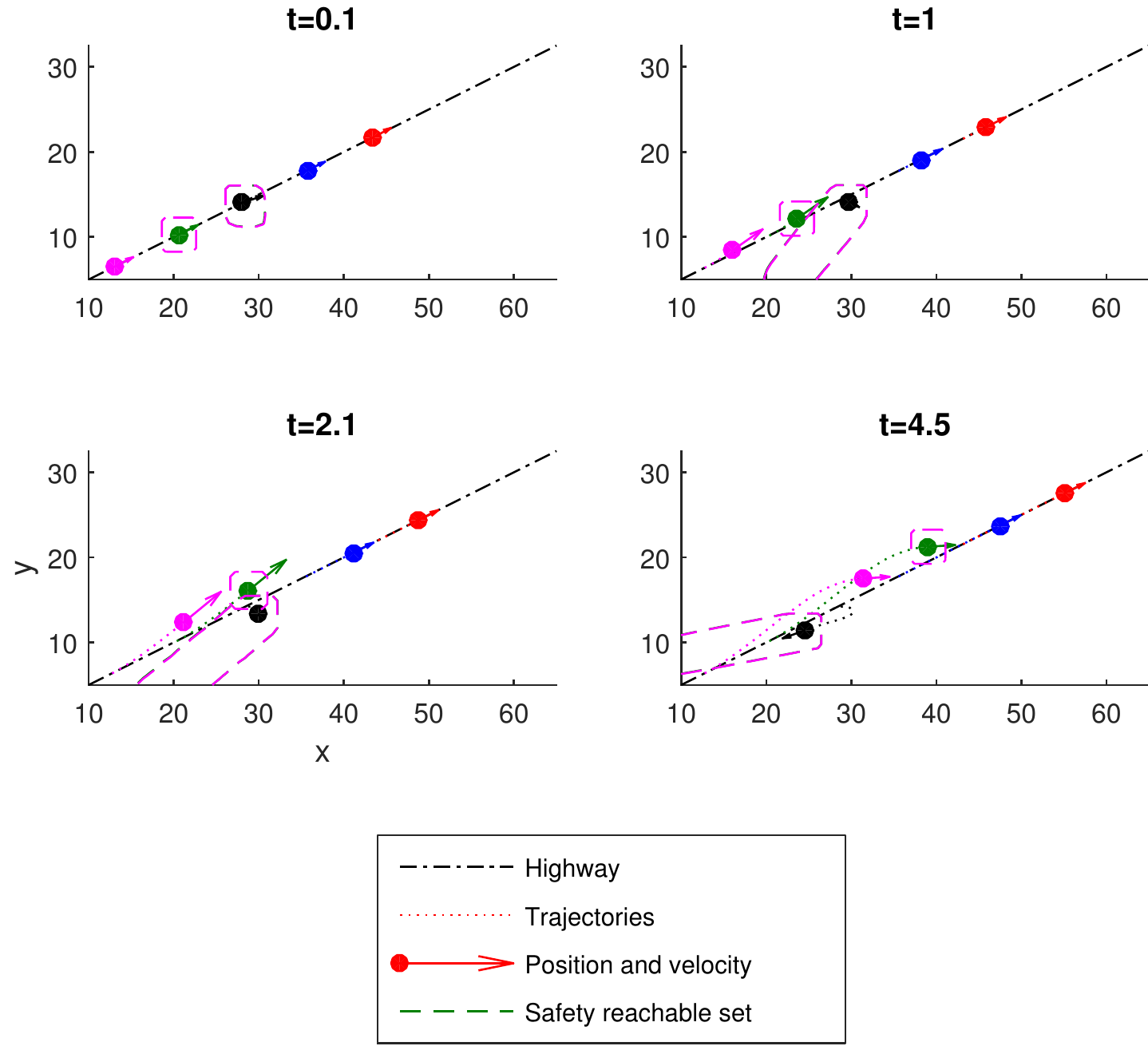}
		\caption{Reachable sets used by trailing quadrotors to avoid colliding with the faulty quadrotor.}
		\label{fig:faulty2}
\end{figure}

Figure \ref{fig:faulty2} shows the platoon of quadrotors, $Q_i,i = 1,...,5$ with $P_i = i$, traveling along the highway. At $t=0, Q_3$ malfunctions and begins to track the highway in reverse. Once $Q_3$ malfunctions, it is removed from the platoon and treated as an intruder. The platoon is then restructured with the faulty quadrotor removed ($Q_{P_i} = Q_{i+1}$ for $i=3,4$). After avoiding $Q_3$, the trailing quadrotors $Q_4$ and $Q_5$ accelerate to reach their new platoon positions. $Q_1$ and $Q_2$ are unaffected by the malfunctioning quadrotor. 

Figure \ref{fig:faulty2} also shows the safe reachable set of $Q_4$ with respect to $Q_3$ (green dashed line), and the safe reachable set of $Q_5$ with respect to $Q_4$ and $Q_3$ (purple dashed lines).

At $t=1,Q_4$ applies the safe controller to avoid entering the safe reachable set with respect to $Q_3$. During $Q_4$'s avoidance maneuver, $Q_5$ simply follows $Q_4$, and does not come across any safety breaches, as shown by the $t=1$ and $t=2.1$ subplots. The safety breach ends soon after, and by $t=4.5$, $Q_4$ begins merging back onto the highway, followed by $Q_5$, in order to continue to follow the platoon. In this particular case, the safety breach is resolved even without any altitude change.

%\begin{figure}
%	\center
%		\includegraphics[width=0.45\textwidth]{fig/faultyNoRsets}
%		\caption{A quadrotor in the platoon becomes faulty. The rest of the platoon safely avoids the quadrotor and continues normal operation as a platoon once the faulty quadrotor has been cleared.}
%		\label{fig:faulty1}
%\end{figure}

\subsection{Intruder Vehicle}
We now consider the scenario in which a platoon of quadrotors encounters an intruder vehicle. To avoid collision, each quadrotor checks for safety with respect to the intruder and any quadrotor in front and behind in the platoon. If necessary, each quadrotor switches to using the safety controller.

\begin{figure}
	\centering
	\includegraphics[width=0.35\textwidth]{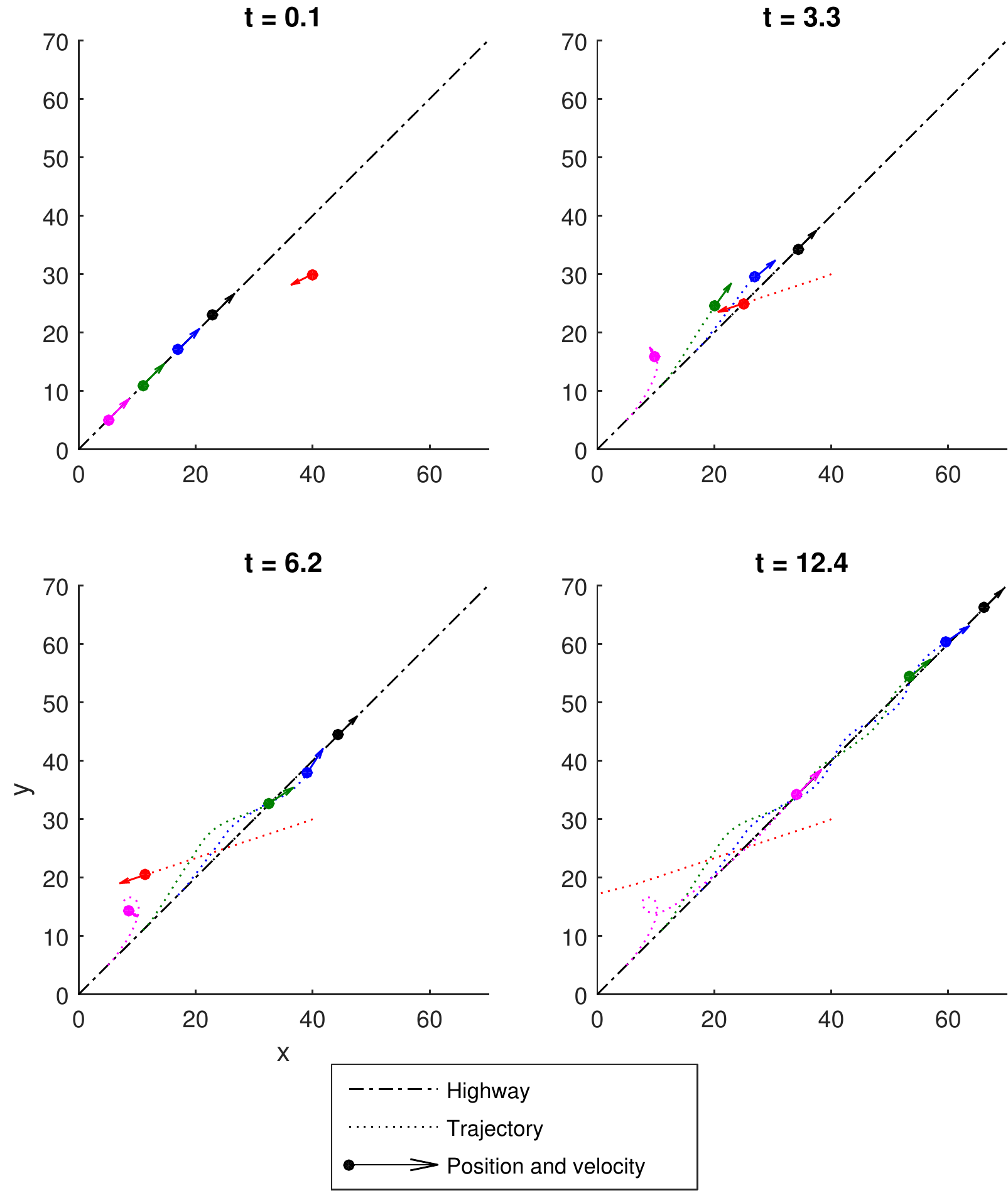}
	\caption{Reaction of a platoon of 4 quadrotors on a highway to an intruder.}
	\label{fig:intruder1}
\end{figure}

Figure \ref{fig:intruder1} shows the simulation result. At $t=0$, a platoon of 4 quadrotors, $Q_{P_i},i=1,\ldots,4$ with $P_i = i$, travel along the highway. An intruder vehicle $Q_0$ (red disk) starts from position $(40,30)$ and heads toward bottom-left of the grid. 

\begin{figure}
	\centering
		\includegraphics[width=0.35\textwidth]{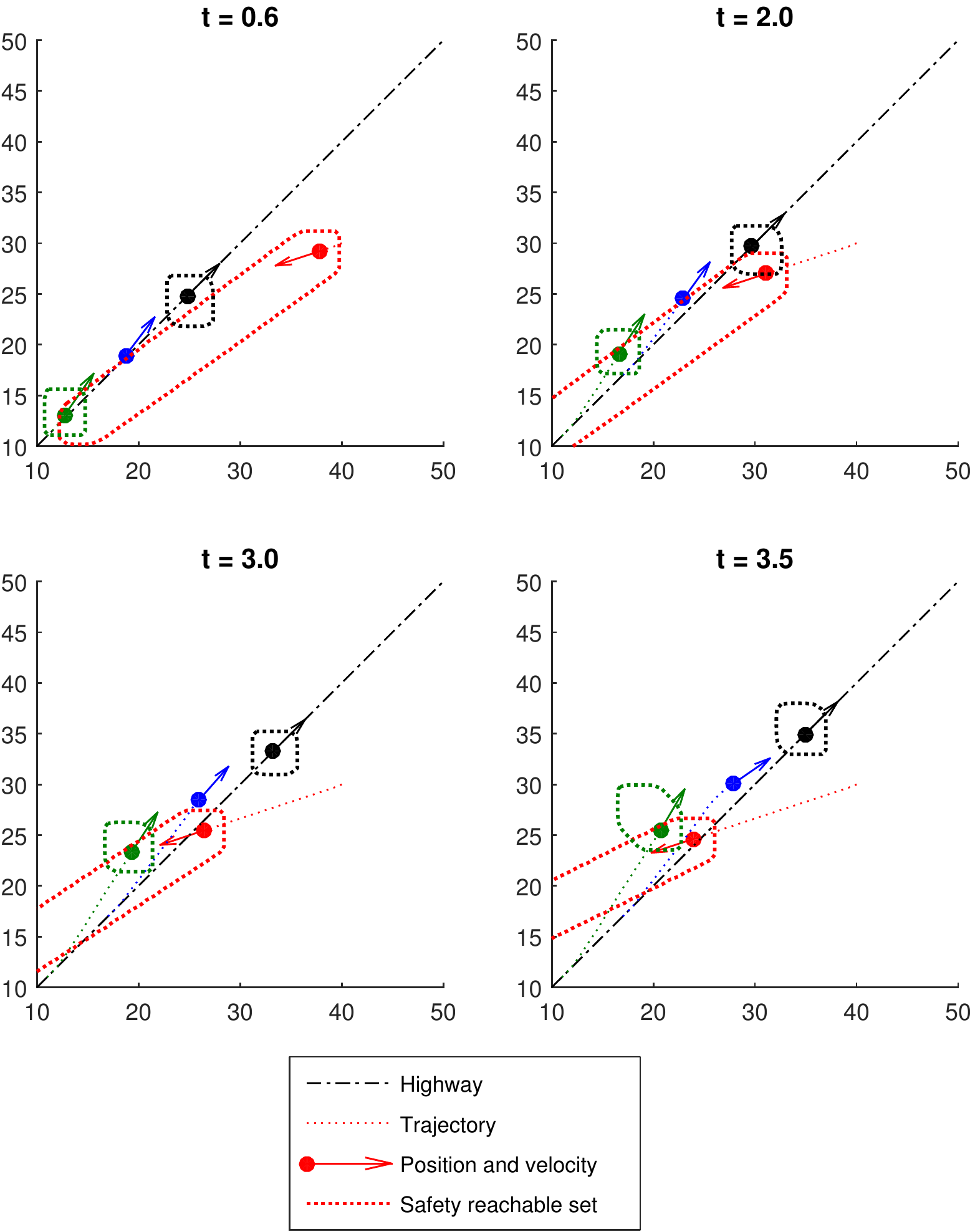}
	\caption{Reachable sets used by quadrotor $Q_2$ to avoid collision with respect to the intruder and quadrotors $Q_1$ and $Q_3$ in front and behind it respectively.}
	\label{fig:intruder2}
\end{figure}

The platoon leader $Q_{P_1}$'s (black disk) safety is unaffected by the intruder. Followers $Q_{P_2}$ (blue disk), $Q_{P_3}$ (green disk) and $Q_{P_4}$ (pink disk), on the other hand, must use the safety controller in order to avoid collision with the intruder ($t=3.3,6.2$). This causes their paths to deviate off the highway. Once each quadrotor is safe relative to the intruder, they rejoin the original platoon ($t=12.4$). Figure \ref{fig:intruder2} illustrates the use of safety reachable sets in this scenario using only $Q_{P_2}$ as an example. The safety reachable sets of $Q_{P_2}$ with respect to the intruder $Q_0$, $Q_{P_1}$ and $Q_{P_3}$ (red, black, green dashed lines) are shown. %With respect to $Q_0$, $Q_{P_2}$'s safety is considered to be breached if $V_S(-t_\text{external},x_{P_2}-x_0) \le 0$. To avoid possible collision with the intruder, $Q_{P_2}$ must remain outside the safety reachable set with respect to the intruder. The same applies to collision avoidance with $Q_{P_1}$ and $Q_{P_3}$. %\textcolor{red}{\sout{Note that safety reachable sets with $Q_{P_1}$ and $Q_{P_3}$ are much smaller than that with the intruder. This is because under normal conditions, vehicles in the same platoon have near-zero relative velocity. For this reason, they may only check possible collisions within the next 1.5 seconds, instead of 3 seconds for vehicles outside of the platoon. This allows vehicles to travel closer together within a platoon and thus increasing throughput on the air highway.}} \textcolor{blue}{The 1.5 seconds is because we made the assumption that malfunctioning vehicles will change altitude within 1.5 seconds.}

Initially, $Q_{P_2}$ ($P_2=2$) is a follower outside all 3 safety reachable sets. At $t=0.6$, $Q_2$ comes to the boundary of the safety set with respect to the intruder and must apply the safety control law to avoid potential future collision. Thus it splits from the original platoon and becomes the leader of a new platoon consisting of itself, $Q_3$ and $Q_4$. $Q_2$ keeps using the safety controller until it is safe with respect to the intruder again at $t=3$. After $t=3$, $Q_2$ is safe to use the liveness controller again to merge back onto the highway and join the original platoon. Note that during the entire time, $Q_2$ maintains safety against the intruder, $Q_1$ and $Q_3$ by always staying outside of all three safety reachable sets.

% Numerical Simuations (1-1.5p)
%% 1: normal operation: form platoon by joining --> just follow --> 1 quadrotor leaves
%% - Mo
%% 2: vehicles already in platoon, one of them goes rogue
%% - Casey
%% 3: vehicles in platoon, intruder vehicle
%% - Qie
%% 4: vehicles in platoon, another vehicle joining, one vehicle in platoon goes rogue
%%
%% followPlatoon - Jaime

% !TEX root = platooning.tex
\section{Conclusions and Future Work}
We considered single-file platoons of UAVs modeled by hybrid systems traveling along air highways. Using HJ reachability, we proposed liveness controllers and built a safety controller around them to ensure no collision can occur from a single safety breach. Additional safety breaches can be handled by multiple altitude ranges in the airspace. Our simulations show that by putting vehicles into single-file platoons, the likelihood of having multiple safety breaches is low, and conflicts involving a single malfunctioning UAV or intruder can be resolved in a single altitude level.

% Conclusion (0.5p)

%%%%%%%%%%%%%%%%%%%%%%%%%%%%%%%%%%%%%%%%%%%%%%%%%%%%%%%%%%%%%%%%%%%%%%%%%%%%%%%%
%\addtolength{\textheight}{1cm}   % This command serves to balance the column lengths

 \bibliographystyle{IEEEtran}
 \bibliography{references}
\end{document}